\documentclass[11 pt]{article}
\usepackage[margin=1in]{geometry}
\usepackage{graphicx}
\usepackage{amsmath}
\usepackage{amsthm}

\usepackage[usenames,dvipsnames]{color}
\definecolor{LightGray}{gray}{0.85}
\definecolor{VeryLightGray}{gray}{0.95}
\usepackage{multirow}
\usepackage{rotating}
\usepackage{colortbl}

\usepackage[numbers,square,comma,sort&compress]{natbib}
\usepackage{ulem}
\usepackage[colorlinks]{hyperref}

\usepackage{lineno}
\usepackage{sectsty}

\theoremstyle{plain}

\providecommand{\Ro}{\mathcal{R}_0}

\newcommand{\beginsupplement}{%
        \setcounter{table}{0}
        \renewcommand{\thetable}{S\arabic{table}}%
        \setcounter{figure}{0}
        \renewcommand{\thefigure}{S\arabic{figure}}%
     }

\newcommand{\captionfonts}{\small}
\makeatletter  % Allow the use of @ in command names
\long\def\@makecaption#1#2{%
  \vskip\abovecaptionskip
  \sbox\@tempboxa{{\captionfonts #1: #2}}%
  \ifdim \wd\@tempboxa >\hsize
    {\captionfonts #1: #2\par}
  \else
    \hbox to\hsize{\hfil\box\@tempboxa\hfil}%
  \fi
  \vskip\belowcaptionskip}
\makeatother   % Cancel the effect of \makeatletter

\allsectionsfont{\sffamily}
  
\newcommand{\comment}[1]{}

\title{\sffamily Modeling Spatial Invasion of Ebola in West Africa}

\author{Jeremy P D'Silva$^{\dagger}$ and Marisa C. Eisenberg$^{\dagger \ddagger}$\\
\small $^\dagger$ Department of Epidemiology, School of Public Health, University of Michigan, Ann Arbor,
\\
\small $^\ddagger$ Department of Mathematics, University of Michigan, Ann Arbor
\\
\small Address correspondence to: marisae@umich.edu.}

\date{}

\begin{document}
\maketitle

\begin{abstract}\sffamily
The 2014-2015 Ebola Virus Disease (EVD) epidemic in West Africa was the largest ever recorded, representing a fundamental shift in Ebola epidemiology with unprecedented spatiotemporal complexity. We developed spatial transmission models using a gravity-model framework to explain spatiotemporal dynamics of EVD in West Africa at both the national and district-level scales, and to compare effectiveness of local interventions (e.g. local quarantine) and long-range interventions (e.g. border-closures). Incorporating spatial interactions, the gravity model successfully captures the multiple waves of epidemic growth observed in Guinea. Model simulations indicate that local-transmission reductions were most effective in Liberia, while long-range transmission was dominant in Sierra Leone. The model indicates the presence of spatial herd protection, wherein intervention in one region has a protective effect on surrounding regions. The district-level intervention analysis indicates the presence of intervention-amplifying regions, which provide above-expected levels of reduction in cases and deaths beyond their borders. The gravity-modeling approach accurately captured the spatial spread patterns of EVD at both country and district levels, and helps to identify the most effective locales for intervention. This model structure and intervention analysis provides information that can be used by public health policymakers to assist planning and response efforts for future epidemics. 
\\
\\
\textbf{Keywords}: Ebola virus disease, transmission modeling, spatial modeling, interventions\\
\textbf{Abbreviations}: EVD, Ebola virus disease; ETU, Ebola treatment unit;, WHO, World Health Organization, IAR, Intervention-amplifying region 
\end{abstract}
%\abbreviations{EVD, Ebola virus disease; ETU, Ebola treatment unit; CI, confidence intervals; ODE, ordinary differential equation; LH, Latin Hypercube}

\section{Introduction}
The outbreak of Ebola virus disease (EVD) in West Africa caused 28,646 cases and 11,323 deaths as of March 30, 2016 \cite{WHO2014_sitrep}. The current outbreak is of the EBOV (Zaire Ebola Virus) strain, the most fatal strain \cite{WHO2014, gire2014genomic}. The largest previous outbreaks of the EBOV strain occurred in the Democratic Republic of the Congo in 1976 and 1995, causing 318 and 315 cases \cite{CDC_chronology}.

Clinical progression includes two broad stages of infection, often characterized as early and late \cite{Ndambi1999}. In the first stage, approximately five to seven days, symptoms include fever, weakness, headache, muscle/joint pain, diarrhea, and nausea \cite{Ndambi1999, CDC_clinicians}. In some patients, the disease progresses to a second stage, with symptoms including hemorrhaging, neurological symptoms, tachypnea, hiccups, and anuria \cite{Ndambi1999,Bwaka1999}. Mortality rates are higher among those exhibiting second-stage symptoms \cite{Ndambi1999, Bwaka1999}. EVD is transmitted through direct contact with an infected individual \cite{Dowell1999}. Transmission risk factors include contact with bodily fluids, close contact with a patient, needle reuse, and contact with cadavers, often prepared for burial by the family \cite{Dowell1999,Roels1999,Epatko2014}.

	This outbreak was primarily in the contiguous countries of Guinea, Sierra Leone, and Liberia, which experienced widespread and intense transmission \cite{WHO2014_sitrep29Apr15,WHO2015_italy, WHO2014_sitrep}. Interventions have included quarantine, case isolation, additional treatment centers, border closures, and lockdowns, restricting travel within a region (as in a military-enforced curfew) \cite{WHO_response}. Public health authorities must allocate resources effectively, focusing personnel and funds to respond best to outbreaks. However, not all response measures are equally beneficial or cost-effective, and testing the relative benefits of each is often impossible or unethical. Modeling thus provides a valuable tool for comparing interventions and identifying areas where interventions are most effective. Modeling analyses can inform public health policy regarding ongoing and future outbreaks.

Dynamic spatial modeling has been proposed as a useful approach to understand the spread of EVD and evaluate response strategies' benefits \cite{Wesolowski2014}. The recent outbreak has sparked an increase in EVD modeling \cite{Yamin2014,Chowell2004,Pandey2014,Rivers2014}. The spread between contiguous countries in the current EVD outbreak highlights the spatial element to its proliferation \cite{WHO2014_sitrep}; however, as previous EVD outbreaks were more localized than the 2014-2015 epidemic \cite{Pourrut2005}, there is little historical data on the geospatial spread of Ebola. Mobility data, which may help inform spatial EVD modeling, is limited, although some studies have highlighted its usefulness and extrapolated based on mobility data from other regions \cite{Halloran2014,Wesolowski2014}.

To address the need for spatial EVD models, models have been developed to examine local spatial spread within Liberia \cite{Merler2015,Wesolowski2014,rizzo2016network} and evaluate the potential risk of international spread using data such as airline traffic patterns\cite{Merler2015,Gomes2014,Ivorra2014}. Our models add to existing spatial models by incorporating intervention comparison and an exploration of the dynamics between different regions, using an uncomplicated model structure. Between-country mobility is important in this epidemic because the borders between countries are porous; borders are drawn across community or tribal lines, resulting in frequent border-crossing to visit family, conduct trade, or settle disputes \cite{Warner2014}. In addition, Santermans and coauthors recently demonstrated that the outbreak is heterogeneous: transmission rates differ between locales \cite{santermans2016spatiotemporal}. Our district-level model offers a mechanistic explanation of observed spatial heterogeneities, adding to the existing literature by explicitly simulating the interactions between districts.

%---------------------------------------------
% Figure: Model Structure & Map
%---------------------------------------------
\begin{figure*}
\centering
\includegraphics[width=\textwidth]{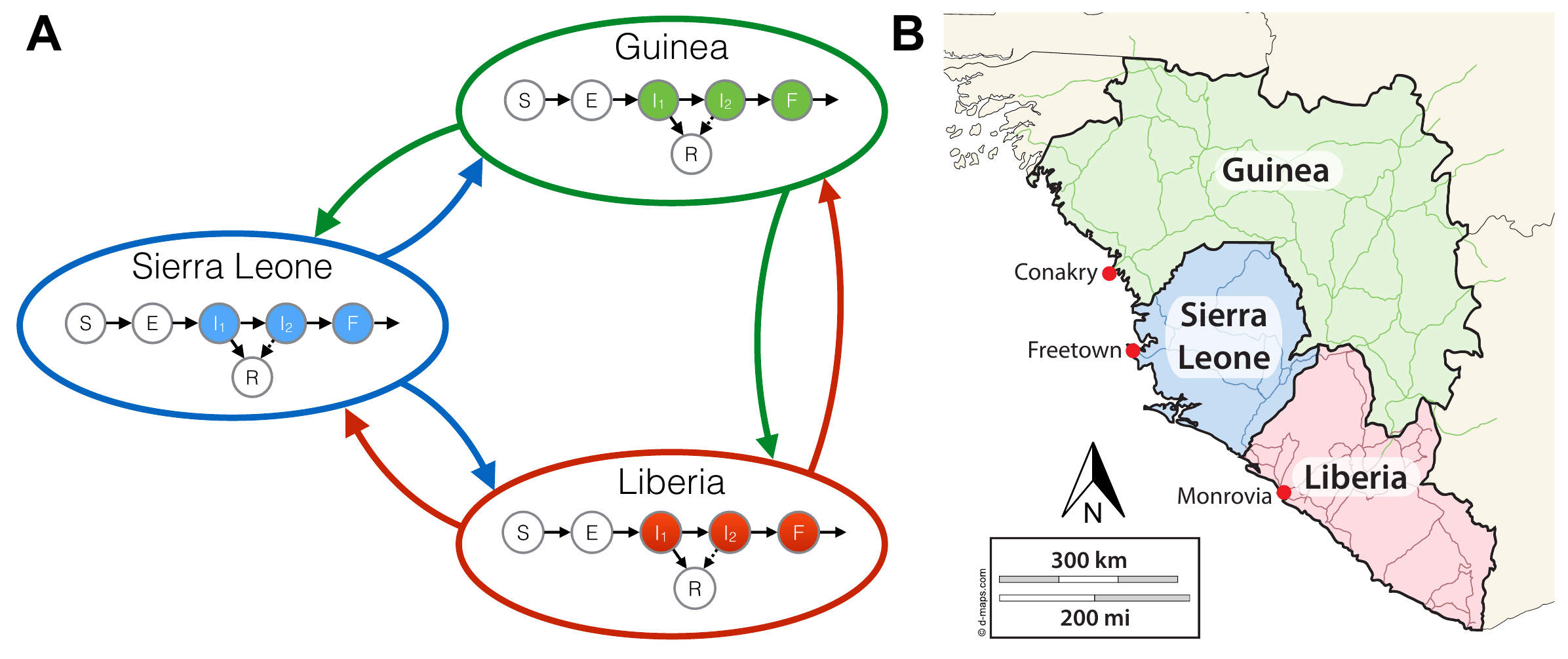}
\caption{A diagram of the model structure (left panel) and a map of West Africa showing the locations of each capital used as a population center for each patch (right panel). In each country, the population is compartmentalized into the following categories: susceptible ($S$), exposed ($E$), infected in stage one ($I_1$), infected in stage two ($I_2$), died but not yet buried ($F$), and recovered ($R$). Transmission is possible from persons in compartments $I_1$, $I_2$, or $F$ of any country, to persons in compartment $S$, either within the country or to the other countries.}
\label{fig:modelmap}
\end{figure*}

In this study, we present spatial models of EVD transmission in West Africa, using a gravity model framework, which captures dynamics of local (within-region) and long-range (inter-region) transmission. Gravity models are used in many population-mobility applications, especially to relate spatial spread of a disease to regional population sizes and distances between population centers \cite{Viboud2006}. Gravity models have been applied to other diseases, including influenza in the US and cholera in Haiti \cite{Tuite2011,Viboud2006}, and have been used to examine general mobility patterns in West Africa\cite{Wesolowski2014}. 

We demonstrate that the gravity modeling approach fits and forecasts case and death trends in each country, and describes transmission between countries. The district-level model successfully simulates local geospatial spread of EVD, indicating that the gravity model framework captures much of the local spatial heterogeneity. Using the model structures, we examined interventions at country and district scales, evaluating the relative success of intervention types as well as the most responsive regions.

\section{Methods}
\subsection{Country-level Model Structure}
	We developed a compartmental gravity model using ordinary differential equations to model spatiotemporal progression of EVD (Figure \ref{fig:modelmap}, model equations and details given in the Supplementary Information). Each country's compartmental structure includes susceptible ($S$), latent ($E$), two-stage infection ($I_1$, $I_2$), funeral ($F$), recovered (R), and deceased (D), based on previous compartmental models \cite{Legrand2007,Eisenberg2015}. Patients in stage $I_1$ can recover or transition to $I_2$, where they may recover with lower probability or transition to $F$, based on clinical observations of symptom progression and mortality \cite{Bwaka1999}. The latent period and two-stage infection are based on the clinical progression of EVD, wherein patients become contagious in $I_1$, with increasing contagiousness in $I_2$ \cite{Ndambi1999,Dowell1999,Yamin2014,Eisenberg2015,chertow2014ebola}. Funerals play a role in transmission because of the high viral load of the deceased and frequent contact with susceptible persons due to cultural burial practices \cite{Khan1999}. The precise relative magnitude of this contribution is unknown; previous models have shown that the relative contributions of each stage are unidentifiable from early data \cite{Eisenberg2015,Weitz2015}. Thus, patients in $F$ were assumed to be as contagious as those in $I_2$.
	
	The spatial component of the model is a three-patch gravity model. Each country is one patch, with a compartmental model within it and the capital as the population center, similar to previous gravity models, which used political capitals/large cities as centers \cite{Viboud2006,Tuite2011}. After EVD appeared in western Guinea, subsequent cases were in the capital, Conakry, suggesting that the capital acts as the central population hub. Similar progression occurred in each country: after EVD cases appeared in border regions, cases soon appeared in the capital  \cite{WHO2015_news}.
	
The force-of-infection term for patch $i$ consists of transmission from within the region and transmission from other patches into patch $i$. Each long-range transmission term is determined by a “gravity” term, proportional to the population sizes, and inversely proportional to the squared distance between them. 

Cumulative local cases were measured by the number of infections in a patch due to local transmission. Cumulative long-range cases were measured by number of cases due to long-range transmission. 

\subsection{District-level spatial model} We developed a more detailed gravity model of West Africa incorporating the 14 districts/areas of Sierra Leone, 15 counties of Liberia, and 34 prefectures of Guinea. This model is structured the same way as the country-level model: a compartmental model in each administrative unit, linked to all other patches by gravity terms. Transmission is separated into local (within-district) and long-range (between-district). 

\subsection{Data and Simulation Setup}
Case and death incidence in each country was collected from World Health Organization (WHO) situation reports on EVD from March 29 to October 31, 2014 \cite{WHO2014_sitrep}. The country-model simulations used data from May 24 through September 30 for fitting model parameters. Data from October was used for validation, to compare model projections to data unused in fitting. Road distance between centers was used for the gravity component \cite{GoogleMaps}. We evaluated direct distance between capitals in the country-level model, yielding similar results. 

For further verification of the model, we also tested the model's ability to fit the outbreak when data was incorporated from March 29. The fits and forecasts from these simulations were similar to those incorporating data from May onward (Supplemental Figures 1-2).

The district model was simulated from March 31, 2014 to January 31, 2015. The model was compared to data on the outbreak’s local geospatial progression from each district, using incidence of cases and deaths from WHO updates and the UN OCHA database \cite{WHO2015_news,UNOCHA}. 

\subsection{Parameter Values and Estimation Methods}
Model parameters were determined from clinical literature and fitting to available data, similarly to previous models \cite{Rivers2014,Eisenberg2015,Lewnard2014}. In the country-level model, nine parameters were fitted to the data: transmission ($\beta$) and death ($\Delta$) as well as three gravity-term constants, $\kappa$, were separately fitted for each country. . The parameter ($\kappa$) adjusts the “gravity” terms to reflect the balance of local and long-range transmission in each country. Parameters, definitions, units, ranges, and sources are given in Supplementary Table S1.  %S\ref{tab:params}. 

In the country-level model, 500 sets of initial values for all parameters were selected with Latin Hypercube (LH) sampling from realistic ranges for parameters, determined from WHO, CDC, and literature data, given in Supp. Table S1 %\ref{tab:params}
 \cite{Ndambi1999, Bwaka1999, Chowell2004, Legrand2007, Eisenberg2015}. For each of the 500 parameter sets, only transmission rates, death rates, and $\kappa$ were fitted by least-squares, using Nelder-Mead optimization in MATLAB \cite{Matlab}. All other parameters were held constant to the values from the LH sample.

In the district-level model, best-fit parameters from the country-level model were used for all parameters except $k_{norm}$.  A different $k_{norm}$ was fitted for each district to reflect differing at-risk populations and reporting rates. Since fitting 63 parameters is highly complex, slow, and possibly unidentifiable, $k_{norm}$ parameters were calculated according to final size, then adjusted by hand to generate curves matching the progression of the outbreak, both overall and in each district. While all other initial conditions were determined from early data and $k_{norm}$, in Conakry, Guinea, the initial conditions were fitted independently from $k_{norm}$, reflecting possible reporting rate differences between initial situation reports and subsequent data. 

\subsection{Intervention Simulations}
Interventions were compared by the number of cumulative cases the model forecasted on October 31 for different types and levels of intervention. Reduction of local transmission represents interventions that limit contact of an EVD patient with others in his or her home country, including quarantine, isolation, hospitalization, and local lockdowns. Reduction of long-range transmission represents border closures, large-scale lockdowns, or other intervention measures that reduce the transmission between countries. 

In the district-level model, interventions were simulated by eliminating the outbreak in one patch (both local and long-range intervention), and simulating to January 31. This is analogous to effective case isolation and quarantine measures within a district, which would limit the transmission of EVD within and from that district. The effectiveness of intervention in one district was measured by calculating the cumulative number of cases reduced in all other districts. The percent reduction from intervention in a district and percent reduction relative to size of the outbreak in that district were calculated.

\section{Results}
\subsection{Country-level Model Fits and Predictions} 
\begin{figure*}
\centering
\includegraphics[width=\textwidth]{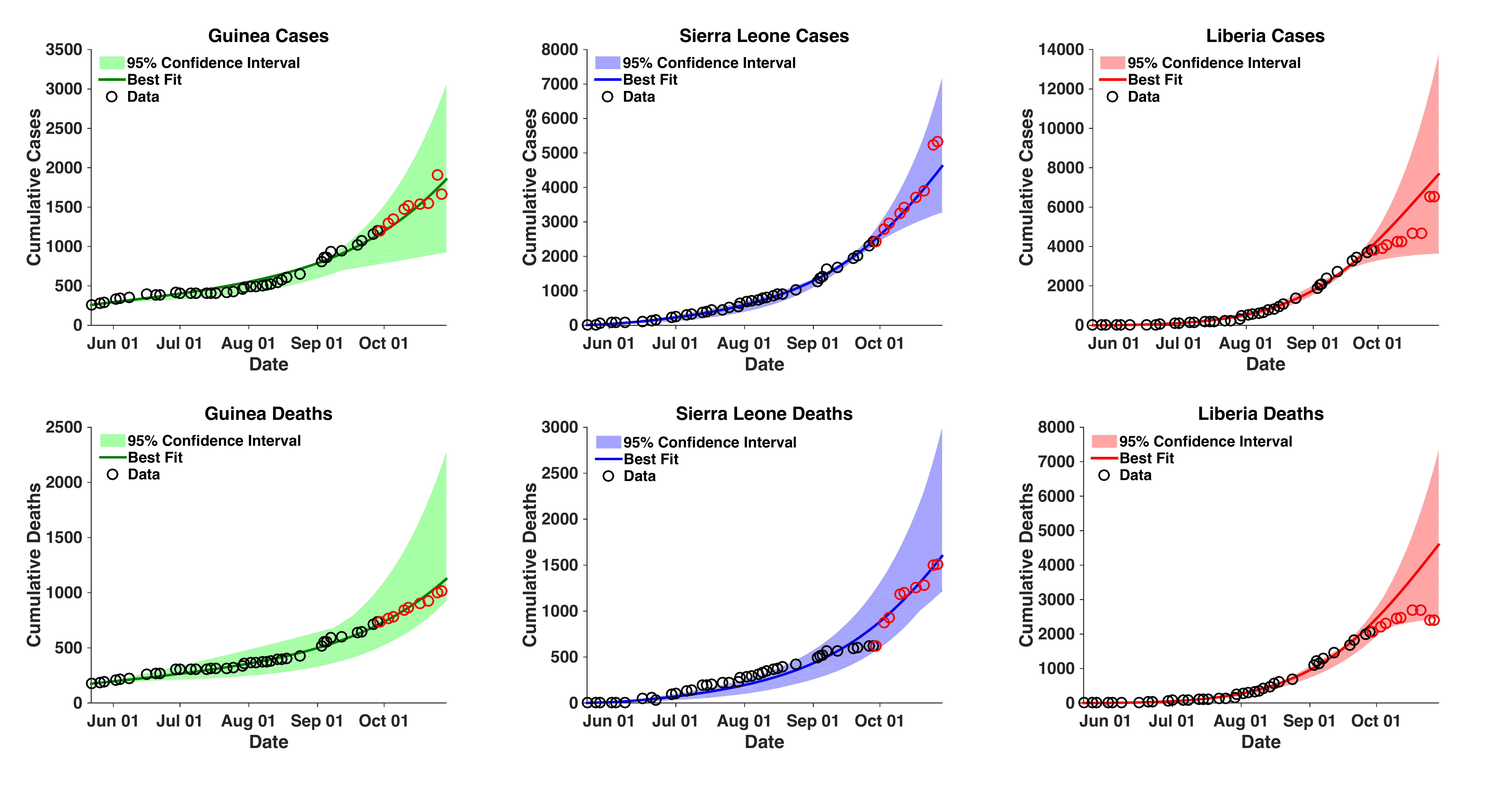}
\caption{Cumulative cases and deaths for each country. Data depicted with black circles (from May 24, 2014 to September 30, 2014) was used for model fitting to the data. Data depicted with red crosses (from October 1 to October 29) is displayed for validation of the model fits. The model projections are from May 24 to October 31, 2014. The black circles represent the data used for fitting; the red crosses represent the data used for validation. The line represents the model’s overall best fit; the shaded regions represent the 95\% confidence interval.}
\label{fig:modelfits}
\end{figure*}

The best-fit accurately fit and forecasted the outbreak data and trends for each country (Figure \ref{fig:modelfits}). The best-fit model forecast of cases on October 29, 2014, was 14,070; the WHO case data on that date was 13,540 \cite{WHO2014_sitrep31Oct14}. 

%---------------------------------------------
% Figure: Country Model Fit/Forecast
%---------------------------------------------

\subsection{Local versus Long-Range Transmission} 
\begin{figure*}
\centering
\includegraphics[width=\textwidth]{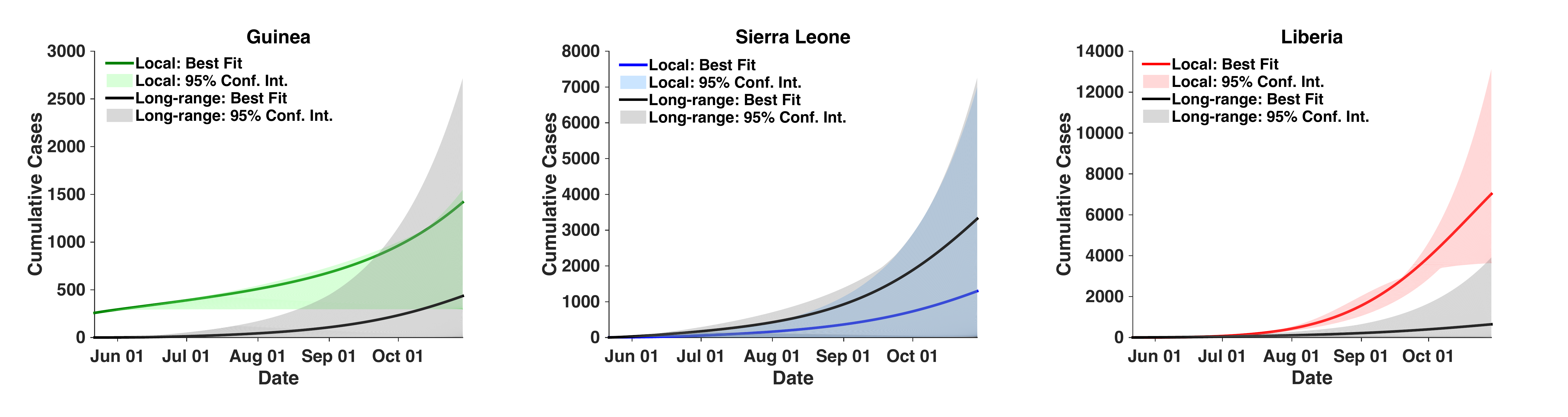}
\caption{Cases from local transmission and long-range transmission. Local transmission, from transmission within the country, is depicted in the color of the country in Figure 1. Long-range transmission, or cases from outside the country, is depicted in gray. Model projections are from May 24 to October 31.}
\label{fig:local_longrange}
\end{figure*}

Cumulative cases in each country due to local and long-range transmission are shown in Figure \ref{fig:local_longrange}. Liberia had more local than long-range transmission. Early on, Guinea had more local transmission, due to initial local cases. As the epidemic progressed, the transmission ranges overlapped, although the best-fit trajectory for long-range transmission remained smaller than the local-transmission contribution. In Sierra Leone, the best-fit trajectory of long-range transmission was more significant than local transmission, although forecasted ranges were similar.

%---------------------------------------------
% Figure: Country Model Local/Long Range
%---------------------------------------------

\subsection{District-level Model}
\begin{figure}
\centering
\includegraphics[width=0.5\textwidth]{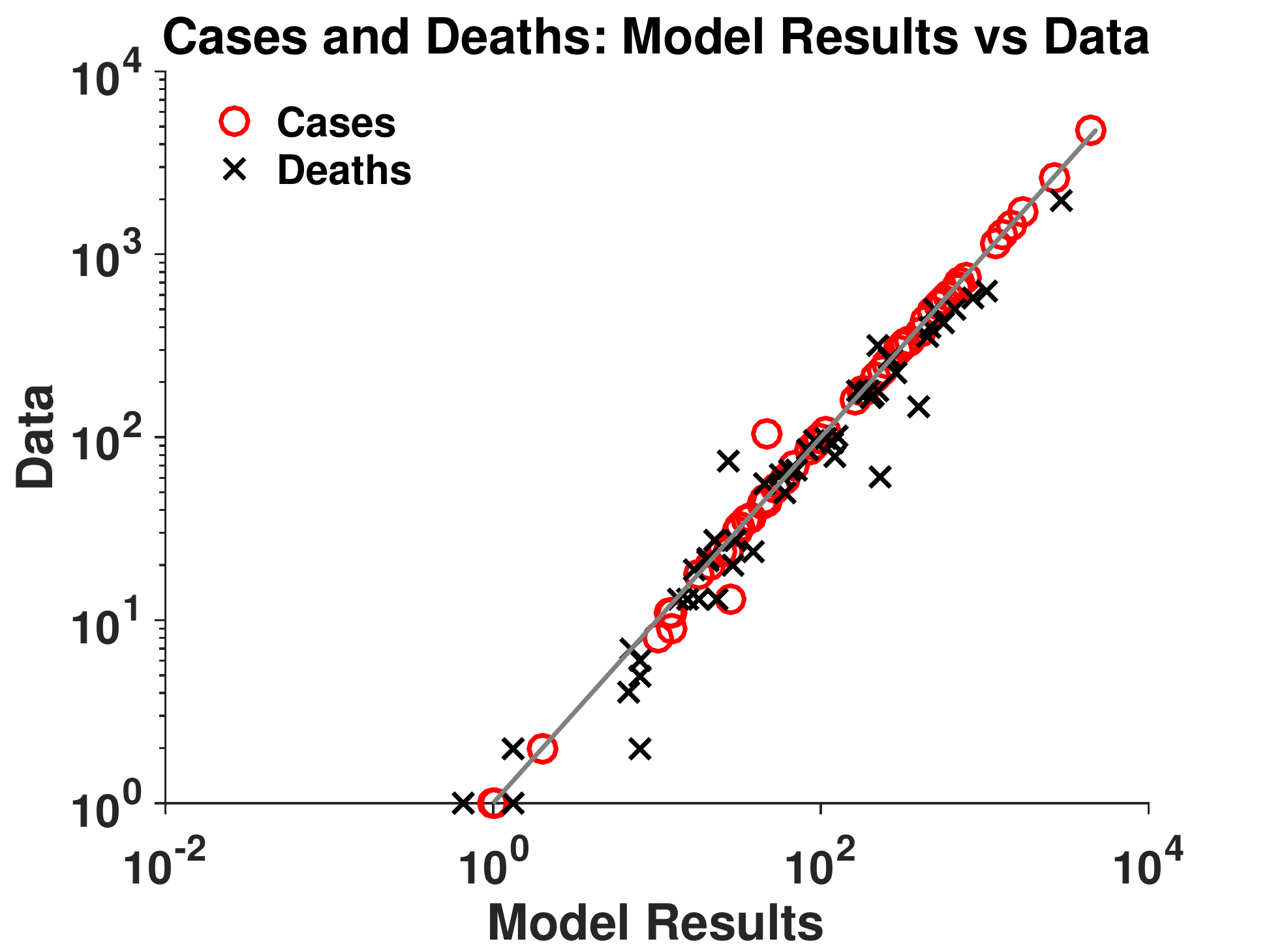}
\caption{The final size of the outbreak in each of the 63 patches: comparison of model values (x axis) and data (y axis). R-squared value of 0.9635.}
\label{fig:districtR2}
\end{figure}

\begin{figure*}
\centering
\includegraphics[width=\textwidth]{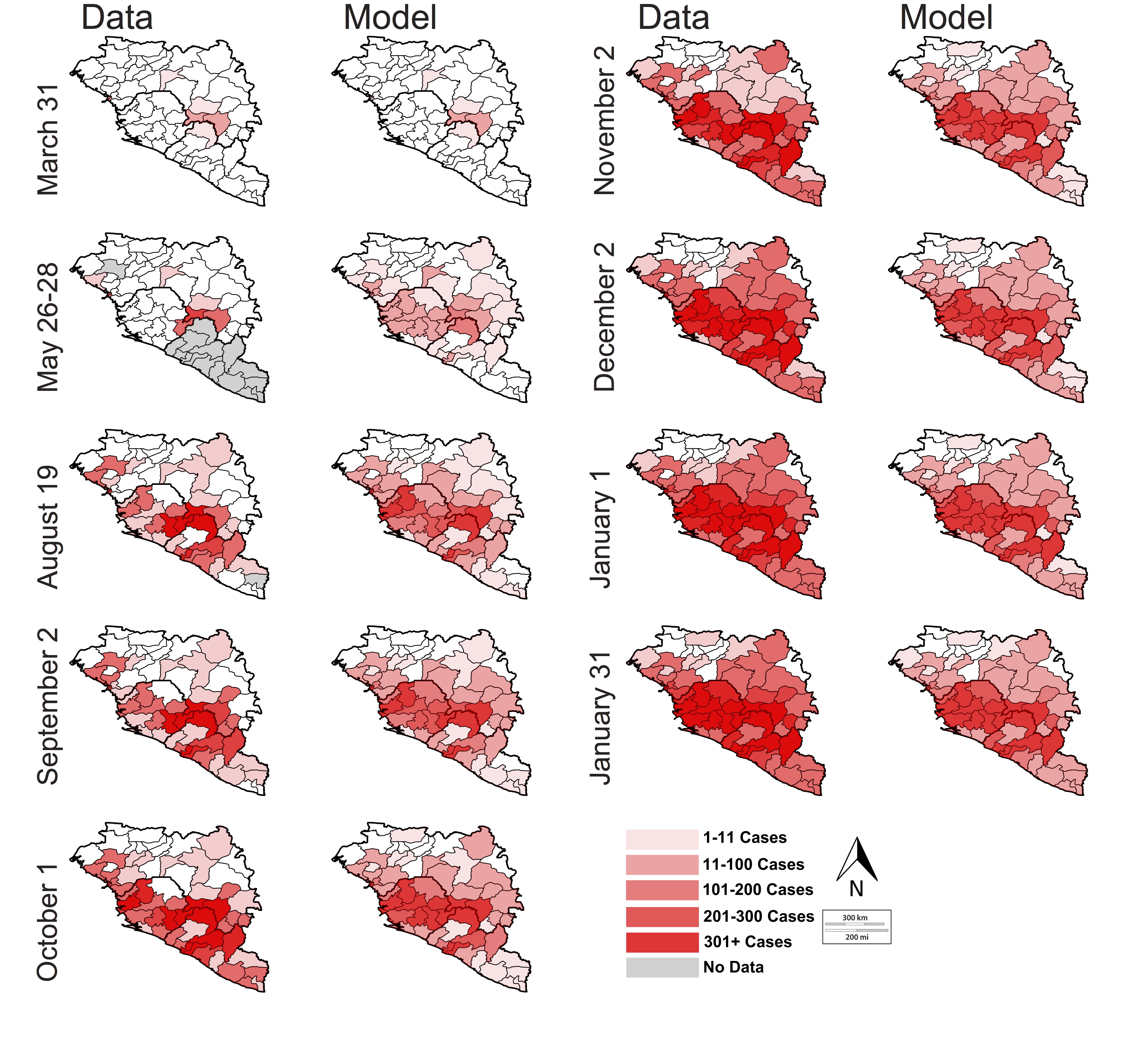}
\caption{Map of geospatial spread of Ebola as determined by the district model (right) and WHO data (left). The shaded regions are the regions with cases of EVD; the intensity represents intensity of transmission (number of cases). Gray regions represent districts for which no data is reported on the given date. On the May 26/28 figure, May 26 is used for Sierra Leone and May 28 is used for Guinea.}
\label{fig:districtMapProg}
\end{figure*}

The district model was successful in fitting outbreak data for cases and deaths in each district. The model captured the final size of the outbreak in each patch with an $R^{2}$ value of 0.96 (Figure \ref{fig:districtR2}) as well as matching the progression of the outbreak (Figure \ref{fig:districtMapProg}). The model captured the temporal dynamics of districts with negligible case counts or high levels of transmission. The time-course plots of all 63 district-level fits are included in the Supplement, as is a scatter plot of all data vs model values ($R^{2}$ = 0.83).

The district-level model accurately captured the trend of spread of EVD through second-level administrative units. While overall patterns at the district level are captured and show the correct ordering, the actual speed of disease spread in the model was faster in some districts than in the data, likely due to a combination of stochastic introductions and reporting delays. The district-level model was able to forecast local spread of EVD, matching data on spatial progression of EVD to different locales as the outbreak intensified (Figure \ref{fig:districtMapProg}). An animation of the district-level simulation is given in the Web Supplement (Video S1).  

\subsection{Intervention Simulations: Country-Level} 
\begin{figure}
\centering
\includegraphics[width=0.6\textwidth]{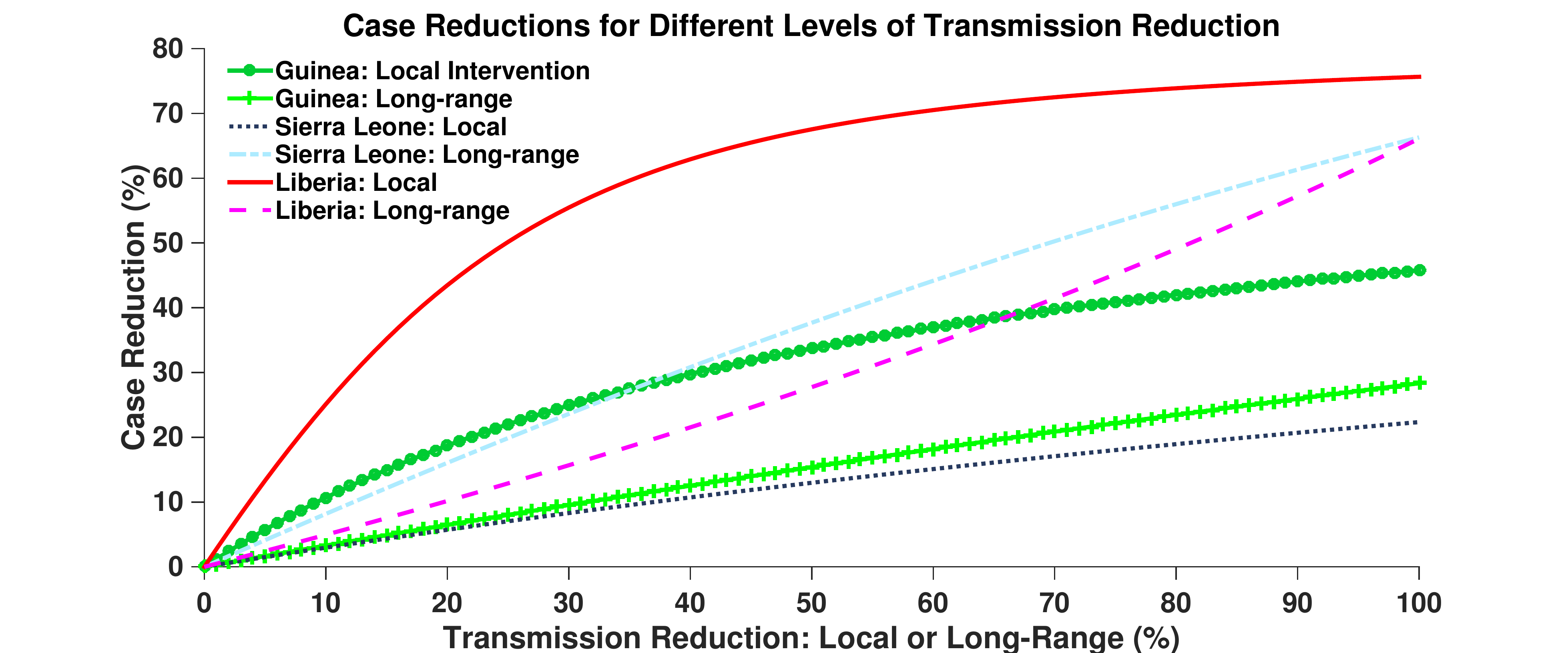}
\caption{Cases reduced by various interventions at the country-level. Transmission reduction represents reduction of local or long-range transmission in increments of one percent. Case reduction is measured by comparing total cases in all 3 countries in the presence of intervention to total cases in the absence of intervention, on October 31, 2014.}
\label{fig:interventions}
\end{figure}

According to our intervention simulations (Figure \ref{fig:interventions}), reduction in local transmission, from interventions such as isolation or improved case-finding and Ebola treatment unit (ETU) capacity, is most effective in Liberia. In the model, eliminating local transmission in Liberia reduced the outbreak by up to 11,000 cases in all countries by October 31, 2014, a 76\% reduction. Reduction of long-range transmission was most effective in Sierra Leone. Eliminating long-range transmission into Sierra Leone reduced the outbreak by up to 9,600 cases in all countries by October 31, a 66\% reduction. 

%---------------------------------------------
% Figure: District Model R^2
%---------------------------------------------

%---------------------------------------------
% Figure: District Model Map Progression
%---------------------------------------------

%---------------------------------------------
% Figure: Country Model Interventions
%---------------------------------------------

\subsection{Intervention Simulations: District-Level} Intervention analysis identified regions in which intervention was most successful (Figure \ref{fig:districtinterventions}). Overall, the most successful regions for intervention were Conakry, Coyah, and Dubreka, Guinea, and Lofa, and Monrovia/Montserrado, Liberia. 

To evaluate whether intervention effects were disproportionately significant compared to cases within the target region, we defined the Intervention-Amplifying Region (IAR) level by the ratio of cases prevented in other patches to cases within the patch, for regions with more than 0.05\% of the epidemic’s cases. Dubreka, Coyah, and Conakry were the most effective IARs (Figure \ref{fig:districtinterventions}).

\section{Discussion}
Our results demonstrate that outbreak dynamics in all three countries can be accurately captured using a gravity-model approach. Moreover, country-level model forecasts successfully predicted cumulative cases and deaths one month ahead (Figure \ref{fig:local_longrange}). The models were able to capture simultaneously the dynamic interactions within and between each region—at both country and district levels.

\subsection{Spatial herd protection: intervention in one region protects others} The model simulations suggest a form of “spatial herd protection”, wherein interventions in one region benefit surrounding regions as well, by reducing the spatial transmission between them. 

In the country-level model, we compared the effects of local and long-range interventions in each country. Local interventions include more strict quarantine and isolation procedures, increased case-finding and ETU capacity, safer burial practices, and other interventions that reduce contact of susceptible persons with infected persons in the local community, as well as behavioral changes that reduce local transmission. The model indicates that these interventions were most effective in Liberia, both in reducing the outbreak in Liberia and mitigating the whole epidemic. Indeed, Liberia’s outbreak had the fastest initial growth rate, it was the first to turn over and end \cite{WHO2014_sitrep}. Our results suggest that Liberian interventions may have had significant indirect effects on the epidemic dynamics in Guinea and Sierra Leone as well. 

The most effective long-range transmission reduction was in Sierra Leone, with 66\% case reduction across all countries when long-range transmission was completely eliminated. This reflects the porous border in Sierra Leone. Based on the district-level intervention analysis, no single district in Sierra Leone acts as a source of long-range transmission to other locales, rather all districts play some combined role. Thus, the country-level impact of long-range transmission is possibly a summative combination of two factors across all districts: long-range transmission into Sierra Leone and early introduction of cases from Sierra Leone into Liberia. 

In the district-level model, the capitals of Montserrado, Liberia and Conakry, Guinea were highly effective intervention sites. Their large population size makes them important in dictating the dynamics of the outbreak in surrounding regions. Dubreka and Coyah, Guinea, near Conakry, were effective intervention sites despite low numbers of cases. This indicates that they influenced the larger outbreaks surrounding them, especially Conakry. Intervention in these districts was especially significant in reducing the scope of the outbreak, not just within the district but also in surrounding regions. 

\subsection{Intervention-amplifying regions: outsized impact on the outbreak dynamics}

\begin{figure}
\centering
\includegraphics[width=\textwidth]{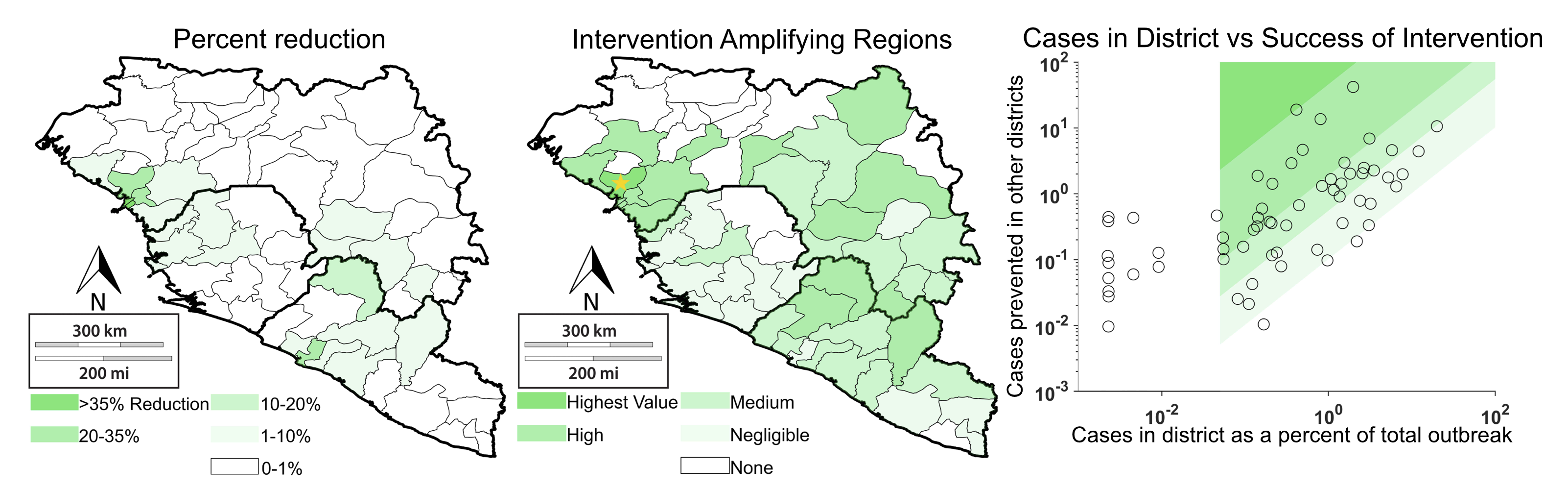}
\caption{Cases reduced by various interventions at the district-level. Transmission reduction consists of elimination of outbreak in one district. Case reduction is measured by (a) percent reduction of outbreak in other 62 patches and (b) percent reduction in other 62 patches relative to size of the outbreak in the intervention site (Intervention-Amplifying Region score). Panel (c) displays the IAR scores on a scatter plot, with the associated quartiles depicted by the shaded regions. The blank region represents the patches with insufficient cases to be considered for intervention.}
\label{fig:districtinterventions}
\end{figure}

In some regions, intervention produced unexpectedly significant reductions in the overall outbreak: these districts amplify the effects of intervention to reduce disproportionately many cases outside the region—these are denoted Intervention-Amplifying Regions (IARs). The most significant IARs include Conakry, Coyah, and Dubreka in Guinea and Montserrado in Liberia (Figure \ref{fig:districtinterventions}). For instance, Dubreka made up only 0.13\% of the outbreak but intervention in Dubreka alone reduced the size of the outbreak in all regions by 21\%, drastically higher than its case contribution would suggest.  Coyah and Dubreka were the most dramatic IARs because they had relatively small numbers of cases. Their proximity to Conakry, which had a large number of cases and also functioned as an IAR, implies that Coyah and Dubreka could have influenced outbreak dynamics in Conakry.

The outsize importance of these districts in the dynamics of the current outbreak makes them a possible target for increased intervention during future outbreaks in West Africa. 

%---------------------------------------------
% Figure: District Model Interventions
%---------------------------------------------

\subsection{Model fits unique case curve in Guinea}
Several papers have noted that Guinea’s peculiar outbreak curve is difficult to fit using simple models, due to plateaus in cumulative incidence between growth periods \cite{Rivers2014,Ivorra2014,Chowell2014_catastrophic}. However, by including spatial interaction between countries, we successfully captured Guinea’s outbreak dynamics (Figure \ref{fig:modelfits}). This suggests that spatial interactions, such as local die-outs of EVD followed by long-range reintroductions, may be responsible for the unusual incidence patterns in Guinea. The country-level model’s ability to capture the singular epidemic progression in Guinea, due to its spatial transmission component, demonstrates that long-range dynamics do not just affect introduction, but also later stages of the outbreak.

\subsection{Discussion of district-level model}
The district-level model, which separates the affected countries into 63 patches based on secondary administrative units, was successful in capturing the overall dynamics of spatial transmission. It captured the spread of EVD from the initial sites of the outbreak to other locales in the region in an order and magnitude similar to the actual progression of the epidemic. This suggests that the county-level model captures the underlying transmission patterns that led to spread of EVD between different locales in the affected countries (Figure \ref{fig:districtMapProg}). These results are perhaps surprising given that the district-level model uses the same transmission parameters for all patches within the same country—this suggests that even though there are likely to be significant local heterogeneities from patch to patch \cite{Chowell2014_exponential}, most of these variations can be captured using the relatively simple framework afforded by the gravity model. 

In fact, a model that incorporates local heterogeneities often requires a large amount of case data at the local scale \cite{chowell2015characterizing}. This gravity model provided a reasonable approximation of the early dynamics of the outbreak, based simply on initial conditions from March 30, soon after surveillance began: it can capture local differences in transmission without complex (difficult-to-simulate) models or extensive datasets. Thus, it could be applicable for early-outbreak forecasting by providing an indication of the areas at greatest risk for Ebola cases, in order to guide resources and aid to those locations.

\subsection{Simplifying Assumptions and Limitations}
Our models focus on early proliferation throughout West Africa: interactions between regions, which led to the spatial spread of Ebola. Regional EVD introductions are inherently stochastic; thus the model did not capture the precise temporal pattern in West Africa in every district. Stochastic model simulations might better characterize the randomness of regional introduction, but a stochastic implementation of the 63-patch regional structure would be computationally demanding. Therefore, a deterministic model offers an elegant method to provide insight into the role of spatial dynamics. 

In addition, we considered the time scales that allowed us to capture the spatial dynamics of interest. In the country-level model, the turnover and re-ignition of the epidemic in Guinea was of particular interest, since simple compartmental models provide little insight into its mechanism \cite{Rivers2014}. Our time-course included the first ignition, turnover, and re-ignition; since simpler compartmental models fail to capture the dynamics of the outbreak, our model’s highly accurate fit implies that the spatial structure captures the dynamics affecting Guinea.

Our time-course ends in November-December 2014, at which time interventions were increasing throughout West Africa. This massive intervention scale-up affected the parameters for the outbreak: in order to simulate the progression of the outbreak after this time, time-variant parameters could be used. The late dynamics were not the focus of this research; however, the model could be implemented with time-variant parameters, potentially providing insight into late dynamics of the outbreak.

\subsection{Future research directions}
While this model captures epidemic dynamics without incorporating explicit movement patterns, further mobility data (e.g. from cell phone carriers) could elucidate dominant movement patterns in West Africa. This data could provide a validation of the gravity model’s success, if the gravity model accounts for these patterns. 

The model does not account for a detailed population structure: an organization of individuals into households, villages, or other communities. The gravity model could be applied with a structured-population (network) model, such as Kiskowski’s model \cite{kiskowski2014three}. This model has been applied to study community-based interventions \cite{kiskowski2015modeling}; it could provide a granular simulation of Ebola dynamics, with possible insights into community and differing levels of local intervention. 

During the course of the outbreak, parameters such as reporting rates, ETU availability, and intervention rates changed. Models that use a broader time-scale than our model see more benefits to incorporating time-variant parameter \cite{rizzo2016network,camacho2015temporal}. In addition, transmission changes as behavior of infected and susceptible individuals changes \cite{rizzo2016network}. Thus, models that incorporate behavioral changes could be successful in capturing that aspect of the epidemic.

\subsection{Conclusions}
We developed dynamic transmission models that account for spatial spread by considering the “gravitational pull” of larger populations and shorter distances in West Africa. The country-level model accurately captures epidemic dynamics and successfully forecasts cases and deaths for all three countries simultaneously. The district-level model captures the progression of EVD between and within districts in West Africa. Our models suggest that reduction of local transmission in Liberia and reduction of long-range transmission in Sierra Leone were the most effective interventions for the outbreak. The models also reveal differences in transmission levels between the three countries, as well as different at-risk population sizes. Our gravity spatial model framework for EVD provides insight into the geographic spread of EVD in West Africa and evaluate the relative effectiveness of interventions on a large, heterogeneous spatial scale. Ultimately, gravity spatial models can be applied by public health officials to understand spatial spread of infectious diseases and guide interventions during disease epidemics.

\section*{Acknowledgments}
This work was supported by the National Institute of General Medical Sciences of the National Institutes of Health under Award Number U01GM110712 (supporting MCE), as part of the Models of Infectious Disease Agent Study (MIDAS) Network. The content is solely the responsibility of the authors and does not necessarily represent the official views of the National Institutes of Health.

\pagebreak

\bibliographystyle{ieeetr}
{\small \sffamily
\bibliography{ebolaPNASrefsJD}
}

\pagebreak
\beginsupplement
\section{Supplementary Information}

\subsection{Model Details}
\noindent\textit{Country-level model}. As discussed above, the model consists of $S$, $E$, $I_{1}$, $I_{2}$, $F$, and $R$ compartments in each country. The model equations are given by:
\begin{equation}
\begin{aligned}
\dot{S}_n &= - (\xi_1 +\xi_2 + \xi_3) S_n\\
\dot{E}_n &= (\xi_1 +\xi_2 + \xi_3) S_n - \alpha E_n\\
\dot{I}_{1n} &=\alpha E_n - \gamma_n I_{1n} - r_{1,n}I_{1n}\\
\dot{I}_{2n} &= \gamma_n I_{1n} - \delta I_{2n} - r_{2,n}I_{2n}\\
\dot{F}_n &= \delta I_{2n} - \delta_2 F_n\\
\dot{R}_n &= r_{1,n}I_{1n} - r_{2,n}I_{2n}\\
\dot{IC}_{1n} &= \alpha E_n\\
\dot{DC}_n &= \delta I_{2n}\\
\xi_1 &= \beta_{1,n} I_{1n} + \beta_{2,n} I_{2n} + \beta_{F,n} F_{n}\\
\xi_2 &= \theta_{n,m} (\beta_{1,n} I_{1m} + \beta_{2,n} I_{2m} + \beta_{F,n} F_{m})\\
\xi_3 &= \theta_{n,l} (\beta_{1,n} I_{1l} + \beta_{2,n} I_{2l} + \beta_{F,n} F_{l})
\end{aligned}
\end{equation}
where $\theta_{n,m} =\kappa_n\frac{\rho_n\rho_m}{(d_{n,m})^\iota}$, $\theta_{n,l} =\kappa_n\frac{\rho_n\rho_l}{(d_{n,l})^\iota}$, and $n$ indicates each patch (Guinea, Sierra Leone, Liberia), and $m$ and $l$ the other two patches (with $m \neq l \neq n$). The risk of new infections from within each patch was represented by the $\xi_1$ term for the “home” patch. TThe risk terms for EVD transmission from outside the patch, $\xi_2$ and $\xi_3$, include $\theta$, the gravity term. The distance exponent of the gravity term, $\iota$, is fixed at two. A range of values for $\iota$ were tested, however we found  that changes in $\iota$ were compensated by changes in the fitted value of $\kappa$. Thus, $\iota$ was fixed for clarity, based on values in previous gravity models \cite{Tuite2011}. There are separate $\kappa$ for each country, so the rate of transmission into each country is different. This reflects the differing border porosities and rates of travel between countries. 
\\

\noindent\textit{District-level model}. Similarly, the district-level model equations for patch $n$ are given by:
\begin{equation}
\begin{aligned}
\dot{S}_n &= - (\xi_1 +\xi_2) S_n\\
\dot{E}_n &= (\xi_1 +\xi_2) S_n - \alpha E_n\\
\dot{I}_{1n} &=\alpha E_n - \gamma_c I_{1n} - r_{1,c}I_{1n}\\
\dot{I}_{2n} &= \gamma_c I_{1n} - \delta I_{2n} - r_{2,c}I_{2n}\\
\dot{F}_n &= \delta I_{2n} - \delta_2 F_n\\
\dot{R}_n &= r_{1,c}I_{1n} - r_{2,c}I_{2n}\\
\dot{IC}_{1n} &= \alpha E_n\\
\dot{DC}_n &= \delta I_{2n}\\
\xi_1 &= \beta_{1,n} I_{1n} + \beta_{2,n} I_{2n} + \beta_{F,n} F_{n}\\
\xi_2 &= \sum\limits_{m=1}^{63} \beta_{1,n} I_{1m} + \beta_{2,n} I_{2m} + \beta_{F,n} F_{m}\\
\theta_{n,m} &=\kappa_n\frac{\rho_n\rho_m}{(d_{n,m})^\iota}
\end{aligned}
\end{equation}

where $\theta_{n,n} := 0$ and $c$ denotes the country to which patch $n$ belongs.

This model structure is similar to that of the country-level model. The main difference is that the long-range transmission is grouped into one $\xi_2$ term, which sums the long-range transmission of every other patch into patch $n$.

\noindent\textit{Model Parameters}.

The model parameters were estimated by simulating total cumulative cases, defined as the integral of the incidence ($\alpha E_n$) multiplied by $k_{norm}$, a correction factor for underreporting and the fraction of the population at risk, among other factors \cite{Eisenberg2015}.

\noindent\textit{Initial Conditions}. 

Initial conditions for both models were determined based on the initial values of the data as follows: the number of susceptible persons in each patch was estimated based on the population size scaled by $k_{norm}$. The total population of each patch was determined using data from the World Bank and national censuses \cite{WorldBank,brinkhoff,liberia,demographie}.
The number of exposed persons was determined as twice the initial number of infected ($I_1$ and $I_2$) persons, based on the $\Ro$ for EVD, which has been estimated to be approximately 2 \cite{Eichner2011,Chowell2004}. The number of infected persons in I1 was determined based on the number of new cases in the previous nine days, based on the incubation period for EVD, and the number of infected persons in I2 was determined by subtracting the number of deaths within the next four days from the number of currently infected persons at the starting date \cite{WHO2014_sitrep}. The number of infected persons from local versus long-range transmission was estimated based on the origin of the outbreak and the location of cases up until the starting point for the data: since the outbreak began in Guinea, the initial cases in Guinea were considered from local transmission, while the initial cases in Sierra Leone and Liberia were considered to be from long-range transmission from Guinea. The number of recovered persons was estimated based on the number of infected persons who did not die within the nine-day period that preceded the starting date. The number of persons in the F class was based on a fraction of the number of persons who died within the period before the starting date, estimated using the burial rate in Table 1. The initial conditions were all scaled to fractions of the population using the parameter  $k_{norm}$.

\begin{figure*}[h]
\centering
\includegraphics[width=\textwidth]{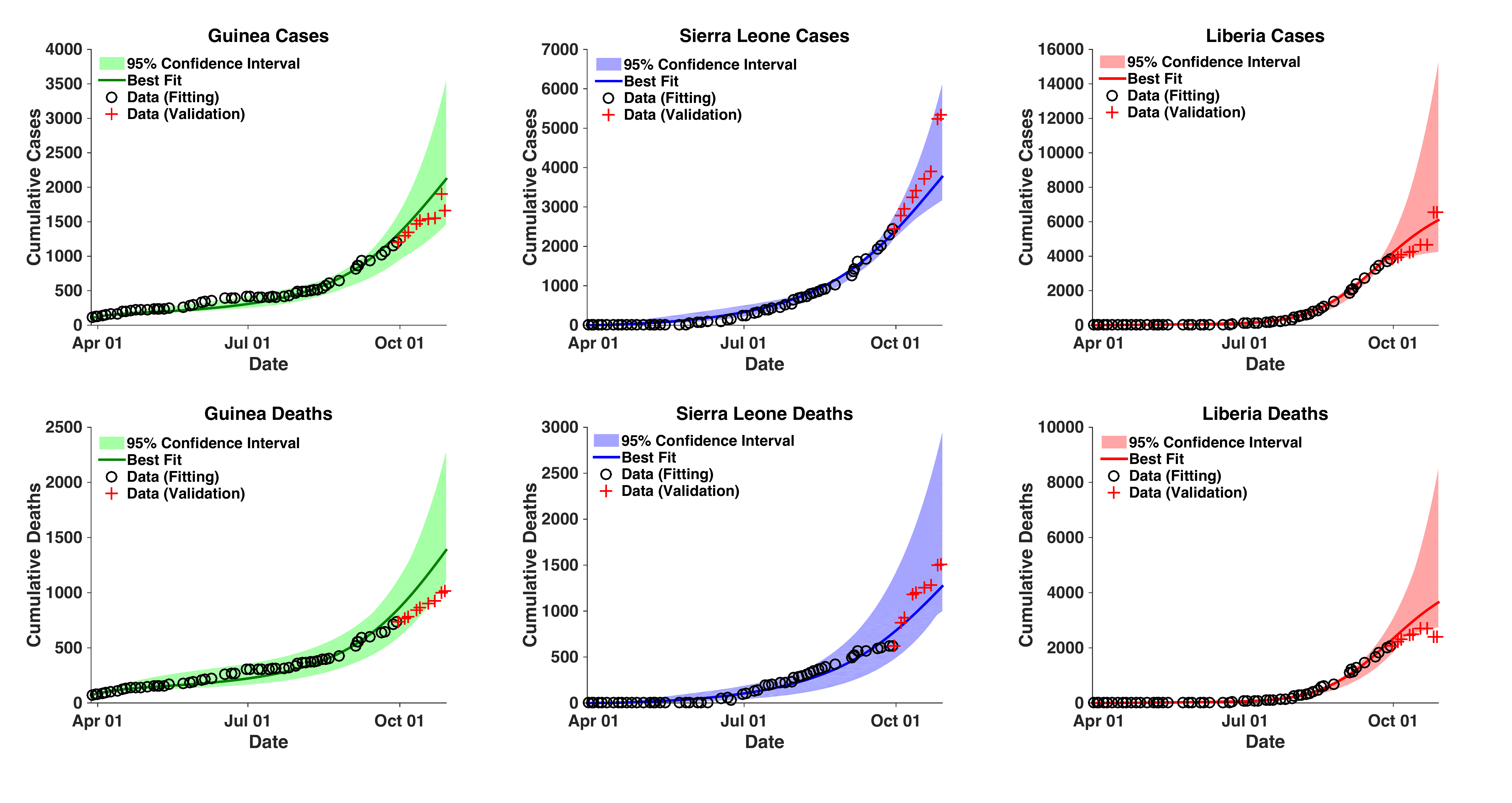}
\caption{Cumulative cases and deaths for each country. Data depicted with black circles (from March 30, 2014 to September 30, 2014) was used for model fitting to the data. Data depicted with red crosses (from October 1 to October 29) is displayed for validation of the model fits. The model projections are from March 30 to October 31, 2014. The black circles represent the data used for fitting; the red crosses represent the data used for validation. The line represents the model’s overall best fit; the shaded regions represent the 95\% confidence interval. (Compare to main text Figure 2.}
\label{fig:modelfits_sup}
\end{figure*}

\begin{figure*}[h]
\centering
\includegraphics[width=\textwidth]{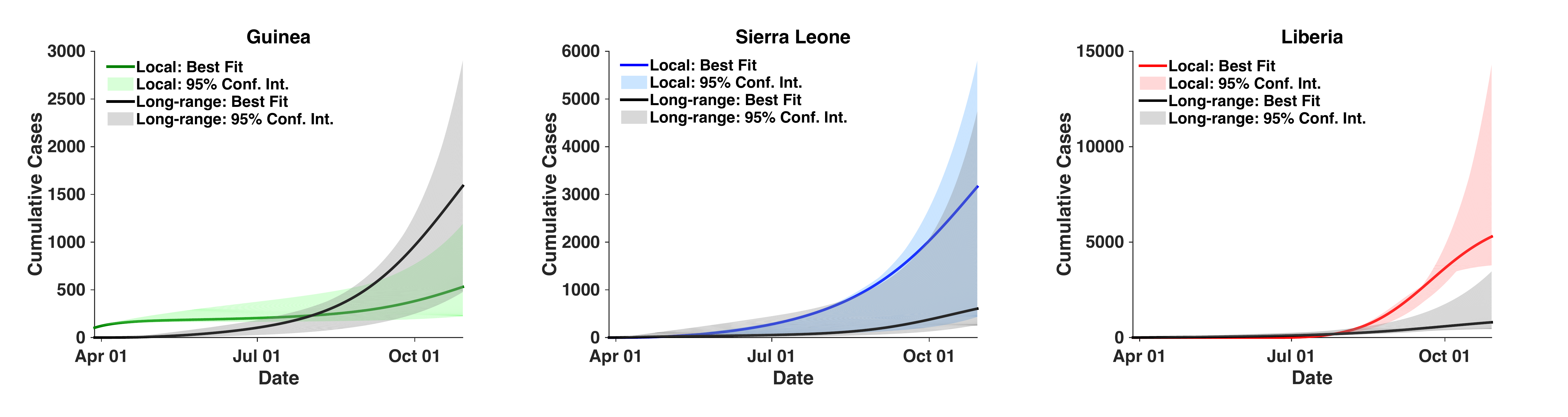}
\caption{Cases from local transmission and long-range transmission. Local transmission, from transmission within the country, is depicted in the color of the country in Figure 1. Long-range transmission, or cases from outside the country, is depicted in gray. Model projections are from March 30 to October 31. (Compare to main text Figure 3.}
\label{fig:local_longrange_sup}
\end{figure*}

We tested the country-level model’s ability to fit the outbreak when data was incorporated using a start date of March 30. In the simulations starting on March 30, local transmission in Liberia was turned off until May (when the first case appeared in Liberia), as the ODE framework used here cannot capture the stochastic process which leads to emergence of an outbreak in a new locale.  The resulting fits and forecasts from these simulations were similar to those incorporating data from May onward (Supplemental Figures \ref{fig:modelfits_sup}, \ref{fig:local_longrange_sup}).

For the district-level model, parameters were determined using sampling from within plausible ranges and from the best-fits of the three-patch models. The β1 transmission rates were determined at the country level, reflecting the differing risks of transmission in the patches in each country. Each $k_{norm}$ was calculated to match final outbreak size, then adjusted by hand. The initial conditions were determined based on data from the WHO reports and from data compiled by the UN. The initial conditions were determined similarly to the country-level model, based on data from the WHO reports and from data compiled by the UN.

The district capitals were considered to be the population centers for all districts except Montserrado, Liberia, and Bonthe, Sierra Leone. Montserrado County is the location of the national capital and largest city, Monrovia, which was considered the population center. The capital of Bonthe is on a coastal island, so Yagoi, the closest city to Bonthe on the coast, was considered the population center. 

In the country-level model, interventions were simulated by reducing transmission parameters for either local or long-range transmission. Local transmission reductions were simulated by reducing ς1, the rate of local infections, in increments of one percent of the original value. Long-range transmission reductions were simulated by reducing ς2 and ς3, the rates of infections from outside the patch, in one-percent increments.

In the district-level model, interventions were simulated by eliminating the outbreak in one district (local transmission, long-range transmission into the district, and long-range transmission out of the district). The effect on the 62 surrounding patches was measured (difference in number of cases in those 62 patches in the normal model simulation versus number of cases in the 62 patches in the intervention simulation). In addition, the IAR score was determined as follows:

\begin{equation}
\begin{aligned}
IAR Score &= \ln{\frac{\% \mbox{ Cases reduced in other districts}}{\% \mbox{ of total outbreak occurring in intervention district}}}
\end{aligned}
\end{equation}

Note that for districts with less than 0.05\% of the outbreak’s cases, the IAR score was defined to be 0. 
\pagebreak
\subsection{Supplementary Figures and Tables}

%---------------------------------------------
% Supp Figure: Country Model Fits
%---------------------------------------------

%---------------------------------------------
% Supp Figure: Local vs. Long-range
%---------------------------------------------

%---------------------------------------------
% Supp Figure: District Model Fits
%---------------------------------------------
\begin{figure*}[h]
\centering
\includegraphics[width=\textwidth]{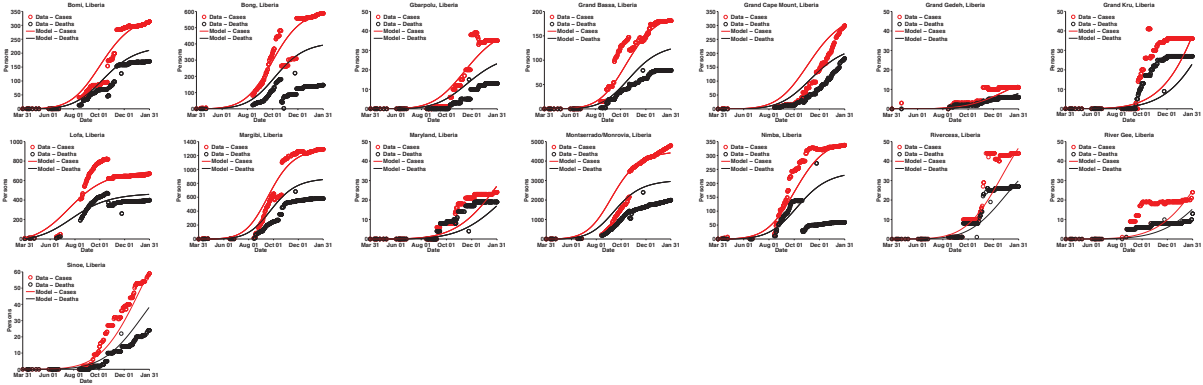}
\caption{Cumulative cases and deaths for each district. Red lines represent cumulative cases (model); black lines represent cumulative deaths (model). Red and black circles represent data for cases and deaths, respectively. Model projections are from March 30, 2014 to January 31, 2015. }
\label{fig:districtfit_sup}
\end{figure*}

%---------------------------------------------
% Supp Figure: District All R^2
%---------------------------------------------
\begin{figure*}[h]
\centering
\includegraphics[width=0.5\textwidth]{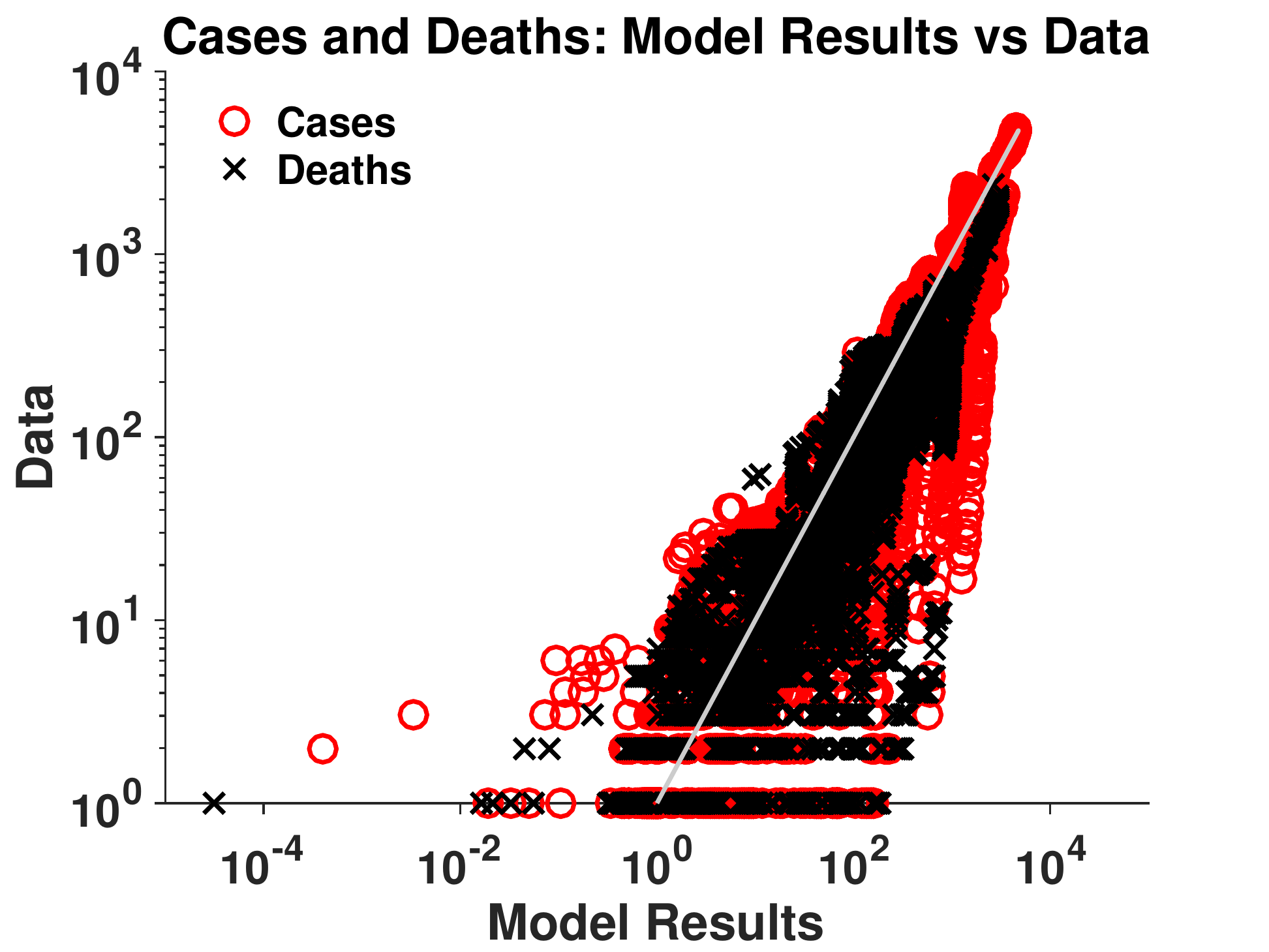}
\caption{Comparison of all data points for cumulative cases and deaths in each of the 63 patches: comparison of model values (x axis) and data (y axis). $R^2$ value of 0.8321.}
\label{fig:districtR2_sup}
\end{figure*}

\newgeometry{margin=0.7in}
\begin{table*}[h]
\centering
\footnotesize
\sffamily \textbf{Table S1}. Parameter values and estimates for the country-level model. 
%\caption{Stuff!}
\begin{tabular}{| c | m{2.1in} m{0.5in} c m{1.6in} m{1in} |}
\hline
\rowcolor{LightGray} Parameter & Definition & Units & Best Fit & Range & Sources\\
\hline
$\alpha$ &Transition rate from $E$ to $I_1$ & days$^{-1}$ & 0.1059 & 0.1-0.125 &\cite{Eichner2011, Ndambi1999} \\

\rowcolor{VeryLightGray}$\beta_{1G}$ &Transmission rate for $I_1$ in Guinea & persons$^{-1}$ days$^{-1}$ & 0.0950 & Initial: 0-0.5, Fitted: 0.0000 to 0.0676 &fitted to data \cite{WHO2014_sitrep}\\

$\beta_{1SL}$ &Transmission rate for $I_1$ in Sierra Leone & persons$^{-1}$ days$^{-1}$&0.0504&Initial: 0-0.5, Fitted: 0.0005 to 0.1858&fitted to data \cite{WHO2014_sitrep}\\

\rowcolor{VeryLightGray}$\beta_{1L}$ &Transmission rate for $I_1$ in Liberia&persons$^{-1}$ days$^{-1}$&0.1555&Initial: 0-0.5, Fitted: 0.0727 to 0.1656&fitted to data \cite{WHO2014_sitrep}\\

$\beta_R$&Ratio of $I_1$ transmission to $I_2$ and $F$ transmission&none&2.4565&1.5-3.0 &\cite{Ndambi1999,Yamin2014}\\

\rowcolor{VeryLightGray}$\beta_{2G}, \beta_{FG}$ &Transmission rate for $I_2$ and $F$ in Guinea&persons$^{-1}$ days$^{-1}$&0.2335&Calculated from $\beta_1$ and $\beta_R$&See $\beta_1$, $\beta_R$\\

$\beta_{2SL}, \beta_{FSL}$&Transmission rate for $I_2$ and $F$ in Sierra Leone&persons$^{-1}$ days$^{-1}$&0.1238&Calculated from $\beta_1$ and $\beta_R$&See $\beta_1$, $\beta_R$\\

\rowcolor{VeryLightGray}$\beta_2L, \beta_{FL}$ &Transmission rate for $I_2$ and F in Liberia&persons$^{-1}$ days$^{-1}$&0.3820&Calculated from $\beta_1$ and &See $\beta_1$, $\beta_R$\\

$\Delta_{1G}$ &Mortality for infected persons in Guinea&none&0.6643&Initial: 0.5-0.9, Fitted: 0.3601 to 1.4023&fitted to data \cite{WHO2014_sitrep}\\

\rowcolor{VeryLightGray}$\Delta_{1SL}$ &Mortality for infected persons in Sierra Leone&none&0.3910&Initial: 0.5-0.9, Fitted: 0.2547 to 0.6159&fitted to data \cite{WHO2014_sitrep}\\

$\Delta_{L}$ &Mortality for infected persons in Liberia&none&0.6710&Initial: 0.5-0.9, Fitted: 0.5453 to 0.8236&fitted to data \cite{WHO2014_sitrep}\\

\rowcolor{VeryLightGray}$\Delta_2$&Fraction of persons in $I_2$ that dies&none&0.9732&0.9-1.0&\cite{Ndambi1999,Bwaka1999}\\

$\delta$&Death rate for persons in $I_2$&days$^{-1}$&0.7597&Calculated from $\Delta_2$ and $k_{02}$&see $\Delta_2$ and $k_{02}$\\

\rowcolor{VeryLightGray}$\delta_2$&Burial rate&days$^{-1}$&0.7425&0.3333-1.0 &\cite{Legrand2007}\\

$k_{02}$&1/duration of $I_2$&days$^{-1}$&0.7806&0.3-0.8 &\cite{Ndambi1999,Khan1999}\\

\rowcolor{VeryLightGray}$k_{21}$&1/duration of $I_1$&days$^{-1}$&0.1822&0.1429 -0.2&\cite{Khan1999,WHO2014}\\

$\iota$&Distance exponent&unitless&2&&fixed\\

\rowcolor{VeryLightGray}$\kappa_1$&Between-patch coupling strength for Guinea&km$^\iota$  persons$^{-2}$&$10^{-10.17}$&Initial: $10^{-8}$-$10^{-12}$,Fitted: $10^{-6.21}$ - $10^{-12.54}$&fitted to data \cite{WHO2014_sitrep}\\

$\kappa_2$&Between-patch coupling strength for Sierra Leone&km$^\iota$ persons$^{-2}$&$10^{-8.27}$&Initial: $10^{-8}$-$10^{-12}$,Fitted: $10^{-6.13}$-$10^{-21.80}$&fitted to data \cite{WHO2014_sitrep}\\

\rowcolor{VeryLightGray}$\kappa_3$&Between-patch coupling strength for Liberia&km$^\iota$ persons$^{-2}$&$10^{-8.62}$&Initial:$10^{-8}$-$10^{-12}$,Fitted: $10^{-7.89}$-$10^{-12.10}$&fitted to data \cite{WHO2014_sitrep}\\

$r_{1G}$ &Recovery rate of persons in $I_1$ in Guinea&days$^{-1}$&0.0578&Calculated from $\Delta_1$, $\Delta_2$, $k_{21}$&fitted to data \cite{WHO2014_sitrep} (See $\Delta_1$, $\Delta_2$, $k_{21}$)\\

\rowcolor{VeryLightGray} $r_{1SL}$ &Recovery rate of persons in $I_1$ in Sierra Leone&days$^{-1}$&0.1090&Calculated from $\Delta_1$, $\Delta_2$, $k_{21}$&fitted to data \cite{WHO2014_sitrep} (See $\Delta_1$, $\Delta_2$, $k_{21}$)\\

$r_{1L}$ &Recovery rate of persons in $I_1$ in Liberia&days$^{-1}$&0.0566&Calculated from $\Delta_1$, $\Delta_2$, $k_{21}$&fitted to data  \cite{WHO2014_sitrep}(See $\Delta_1$, $\Delta_2$, $k_{21}$)\\

\rowcolor{VeryLightGray} $r_2$ &Recovery rate of persons in $I_2$ &days$^{-1}$&0.0209&Calculated from $\Delta_2$, $k_{02}$&\cite{Ndambi1999,Bwaka1999}\\

$\gamma_G$&Rate and proportion of movement of persons from $I_1$ to $I_2$ in Guinea &days$^{-1}$&0.1243&Calculated from $\Delta_1$, $\Delta_2$, $k_{21}$&fitted to data \cite{WHO2014_sitrep} (See $\Delta_1$, $\Delta_2$, $k_{21}$)\\

\rowcolor{VeryLightGray}$\gamma_{SL}$ &Rate and proportion of movement of persons from $I_1$ to $I_2$ in Sierra Leone&days$^{-1}$&0.0732&Calculated from $\Delta_1$, $\Delta_2$, $k_{21}$&fitted to data  \cite{WHO2014_sitrep}(See $\Delta_1$, $\Delta_2$, $k_{21}$)\\

$\gamma_L$&Rate and proportion of movement of persons from $I_1$ to $I_1$ in Liberia&days$^{-1}$&0.1256&Calculated from $\Delta_1$, $\Delta_2$, $k_{21}$&fitted to data  \cite{WHO2014_sitrep}(See $\Delta_1$, $\Delta_2$, $k_{21}$)\\

\rowcolor{VeryLightGray}$\xi_1$ &Transmission rate from S to $I_1$ for locally transmitted cases of EVD&days$^{-1}$&Varies&Calculated from $\beta_1$, $\beta_2, \beta_F$, $I_1$, $I_1$, $F$&See $\beta_1$, $\beta_2, \beta_F$\\

$\xi_2, \xi_3$ &Transmission rate from S to $I_1$ for long-range transmitted cases of EVD&days$^{-1}$&Varies&Calculated from $\beta_1$, $\beta_2, \beta_F$, $I_1$, $I_1$, $F$&See $\beta_1$, $\beta_2, \beta_F$\\

\rowcolor{VeryLightGray}$\theta$&Influence of long range cases on home patch&unitless& -- &Calculated from $\rho$, $d$, $\iota$, $\kappa$&See $\rho$, $d$, $\iota$, $\kappa$ \\

$P_n$ &Population in patch $n$&persons&--&Fixed for each country&\cite{WorldBank}\\

\rowcolor{VeryLightGray}$d_{n,m}$ &distance between patch $n$ and $m$ &km &--&fixed for each combination of patches&\cite{GoogleMaps}\\

$k_{norm}$ &Correction factor for reporting rate, population at risk, and other factors &unitless&0.0030&0.001-0.1&Sampled\\
\hline
\end{tabular}
\label{tab:params}
\end{table*}

\begin{table*}[h]
\centering
\def\arraystretch{1.2}
%\small
\sffamily \textbf{Table S2}. Parameter values for the district-level model. 
%\caption{Stuff!}
\begin{tabular}{| c | m{2.1in} c m{1in} m{1in} |}
\hline
\rowcolor{LightGray} Parameter & Definition & Units & Value & Sources\\
\hline
$\alpha$ &Transition rate from $E$ to $I_1$ & days$^{-1}$ & 0.1059 &\cite{Eichner2011, Ndambi1999} \\

\rowcolor{VeryLightGray}$\beta_{1G}$ &Transmission rate for $I_1$ in Guinea & persons$^{-1}$ days$^{-1}$ & 0.0950 &  fitted to data \cite{WHO2014_sitrep}\\

$\beta_{1SL}$ &Transmission rate for $I_1$ in Sierra Leone & persons$^{-1}$ days$^{-1}$&0.0504 &fitted to data \cite{WHO2014_sitrep}\\

\rowcolor{VeryLightGray}$\beta_{1L}$ &Transmission rate for $I_1$ in Liberia&persons$^{-1}$ days$^{-1}$&0.1555&fitted to data \cite{WHO2014_sitrep}\\

$\beta_R$&Ratio of $I_1$ transmission to $I_2$ and $F$ transmission&none&2&\cite{Ndambi1999,Yamin2014}\\

\rowcolor{VeryLightGray}$\Delta_{1G}$ &Mortality for infected persons in Guinea&none&0.6643&fitted to data \cite{WHO2014_sitrep}\\

$\Delta_{1SL}$ &Mortality for infected persons in Sierra Leone&none&0.3910&fitted to data \cite{WHO2014_sitrep}\\

\rowcolor{VeryLightGray}$\Delta_{L}$ &Mortality for infected persons in Liberia&none&0.6710&fitted to data \cite{WHO2014_sitrep}\\

$\Delta_2$&Fraction of persons in $I_2$ that dies&none&0.97&\cite{Ndambi1999,Bwaka1999}\\

\rowcolor{VeryLightGray}$\delta_2$&Burial rate&days$^{-1}$&0.9 &\cite{Legrand2007}\\

$k_{02}$&1/duration of $I_2$&days$^{-1}$&0.8 &\cite{Ndambi1999,Khan1999}\\

\rowcolor{VeryLightGray}$k_{21}$&1/duration of $I_1$&days$^{-1}$&0.2&\cite{Khan1999,WHO2014}\\

$\gamma_G$&Rate and proportion of movement of persons from $I_1$ to $I_2$ in Guinea &days$^{-1}$&0.1370&fitted to data \cite{WHO2014_sitrep} (See $\Delta_1$, $\Delta_2$, $k_{21}$)\\

\rowcolor{VeryLightGray}$\gamma_{SL}$ &Rate and proportion of movement of persons from $I_1$ to $I_2$ in Sierra Leone&days$^{-1}$&0.0806&fitted to data  \cite{WHO2014_sitrep}(See $\Delta_1$, $\Delta_2$, $k_{21}$)\\

$\gamma_L$&Rate and proportion of movement of persons from $I_1$ to $I_1$ in Liberia&days$^{-1}$&0.1384&fitted to data  \cite{WHO2014_sitrep}(See $\Delta_1$, $\Delta_2$, $k_{21}$)\\

\rowcolor{VeryLightGray}$\iota$&Distance exponent&unitless&2 &fixed\\

$\kappa$&Between-patch coupling strength &km$^\iota$  persons$^{-2}$& $10^{-8.5}$ &fitted to data \cite{WHO2014_sitrep}\\

\rowcolor{VeryLightGray}$\theta$&Influence of long range cases on home patch&unitless& Calculated from $\rho$, $d$, $\iota$, $\kappa$&See $\rho$, $d$, $\iota$, $\kappa$ \\

$P_n$ &Population in patch $n$&persons&Fixed for each patch&\cite{WorldBank}\\

\rowcolor{VeryLightGray}$d_{n,m}$ &distance between patch $n$ and $m$ &km & fixed for each combination of patches&\cite{GoogleMaps}\\

$k_{norm}$ &Correction factor for reporting rate, population at risk, and other factors &unitless&Varies by patch&Fitted to data\\

\hline
\end{tabular}
\label{tab:params}
\end{table*}

\restoregeometry

\clearpage

%---------------------------------------------
% Video: Spatial spread
%---------------------------------------------
{\noindent Caption for Supplementary Animation: \\
\noindent  \textbf{Video S1.} A GIF representing the progression of Ebola according to the model from March 30, 2014, to January 31, 2015. Color intensity of district represents number of cumulative cases in that district: darker color represents higher number of cases.}

\end{document}